\documentclass[11pt,twoside
]{article}
\usepackage{baaa2008}
\usepackage{graphicx}
\usepackage{subfigure}
\usepackage{psfrag}
\usepackage{amssymb}
\usepackage[spanish,activeacute,english]{babel}
\usepackage[latin1]{inputenc}
\usepackage[T1]{fontenc} 
\usepackage{ae,aecompl} 
\usepackage{latexsym}
\usepackage{verbatim}
\usepackage{amsmath}
\usepackage{amsfonts}
\usepackage{amssymb}
\usepackage{wasysym}
\usepackage[colorlinks=true,dvips]{hyperref}

\begin{document}
\myselectenglish
\vskip 1.0cm
\markboth{F. Faifer et al.}%
{GEMINI-GMOS spectroscopy in the Antlia cluster}

\pagestyle{myheadings}
\vspace*{0.5cm}
\parindent 0pt{PRESENTACI\'ON ORAL}
\vskip 0.3cm
\title{GEMINI-GMOS spectroscopy in the Antlia cluster}
\author{F. Faifer$^{1,2}$, A. Smith Castelli$^{1,2}$,  L. P. Bassino $^{1,2}$, T. Richtler$^{3}$, S. A. Cellone$^{1,2}$}
\affil{
(1) Facultad de Ciencias Astron\'omicas y Geof\'isicas (FCAG, UNLP)\\
(2) Instituto de Astrof\'isica de La Plata (CCT La Plata, CONICET - UNLP) and CONICET, Argentina\\
(3) Universidad de Concepci\'on, Chile\\
}
\begin{abstract} We present preliminary results of a spectroscopic study 
performed in the Antlia cluster through GEMINI-GMOS data. We derived new 
radial velocities that allow us to confirm the cluster membership 
of several new faint galaxies, as well as to identify very interesting
background objects.  
\end{abstract}
\begin{resumen}
Presentamos resultados preliminares de un estudio espectrosc\'opico de la
poblaci\'on de galaxias del c\'umulo de Antlia, en base a datos
obtenidos con GEMINI-GMOS. Hemos determinado nuevas velocidades radiales 
las cuales permiten confirmar la pertenencia al c\'umulo de nuevas galaxias 
d\'ebiles, as\'i como la identificaci\'on de objetos de fondo con caracter\'isticas 
interesantes.
\end{resumen}

\section{Introduction}
The Antlia cluster is the nearest populous galaxy cluster after Fornax and 
Virgo (d = 35.2 Mpc, Dirsch et al. 2003). Ferguson \& Sandage (1990, hereafter 
FS90) estimate the number of Antlia galaxies to be about 420, while its 
central galaxy density is claimed to be a factor 1.4 higher
than in Fornax, and almost a factor of 2 higher than in Virgo. 

Antlia exhibits a complex structure consisting of several subgroups, the most 
important ones being dominated by the giant ellipticals NGC 3258 and NGC 3268. 
X-ray studies found extended emission around both subgroups (Pedersen et al. 
1997, Nakazawa et al. 2000), and our studies of the globular cluster systems 
around both dominant galaxies (Dirsch et al. 2003, Bassino et al. 2008)
point to possible interaction processes between them. The overall structure 
suggests an ongoing merger of several subclusters and a considerable structural
depth, and perhaps a connection with the Hydra-Centaurus supercluster 
(Hopp \& Materne 1985). 

In spite of its proximity, high density and interesting substructure, the 
Antlia cluster had been scarcely investigated in the optical and 
spectroscopically until we started with our {\it Antlia Project}. Within this
project, we are carrying out a photometric study of the galaxy 
population of this cluster (Smith Castelli et al. 2008a,b, Bassino 
et al. 2008) based on wide-field MOSAIC data from CTIO, in $C$ and $T_1$ 
(Washington photometric system), and VLT FORS1 images in $V$ and $I$. 
Integrated magnitudes and colours have been obtained for almost 100 galaxies 
listed in the FS90 catalogue (Smith Castelli et al. 2008a). Early-type 
galaxies follow a tight $T_1$ versus $(C-T_1)$ colour-magnitude relation (CMR) 
($\sigma_{(C-T_1)}\sim 0.07$ mag), that spans 9 mag in luminosity with no 
apparent change of slope. In addition, we have identified about 30 new dwarf 
galaxy candidates (ellipticals, spheroidals, irregulars) by means of 
morphology and color criteria, thus extending the FS90 photographic survey 
from $B_\mathrm{T} \sim 18$ mag to a fainter magnitude limit $B_\mathrm{T} =
22.6$ mag (Smith Castelli et al. in preparation).

In this contribution, we present preliminary results from a GEMINI-GMOS
(semester 2008A) spectroscopic study of faint galaxies in the Antlia
cluster. NED contains, within our central 
field, only 30 radial velocities of bright Antlia members. Thus, the immediate 
outcome of these data will be radial velocities for our target galaxies. Since 
these would be (excluding the bright galaxies) the second set of spectroscopic 
data for Antlia (the first one was obtained with MAGELLAN-IMACS, Smith 
Castelli et al. 2008a), they will primarily answer basic questions regarding 
galaxy membership versus background galaxies. The recently obtained  
radial velocities for several Antlia FS90 dwarfs as well as for new dwarf 
candidates have allowed us to extend the CMR defined by spectroscopically 
confirmed early-type members by 4 mag.

\vskip 0.5cm

\section{The data}

We obtained multi-object spectra in three fields placed in the central region 
of Antlia. The fields were chosen so that they contained the largest number
of galaxy candidates as possible (Fig.\,1), with the candidates selected 
from the FS90 Antlia Group catalogue, and from previous morphological studies. 
We placed a mean of 20 slits on each field. Out of them, about 10 were used for 
galaxy candidates and templates (bonafide Antlia members), and we included more 
slits for ultra-compact dwarf (UCD) and globular cluster candidates located in the 
same fields.

The spectroscopic data were obtained in 2008 February 14 and March 15-16 with 
the Gemini Multi-Object Spectrograph (GMOS) mounted on the Gemini-South 
telescope (program ID: GS-2008A-Q-56, PI: T. Richtler). The B600\_G5303 
grating blazed at 5000\,\AA  was used, with three different central 
wavelengths (5000, 5050 and 5100\,\AA) in order to fill in the CCD gaps. A slit 
width of 1 arcsec was used in average seeing of 0.5 - 0.6 arcsec. This 
configuration gives a wavelength coverage of 3500 - 7200\,\AA  depending on 
the positions of the slits, and a resolution (fwhm) of $\sim 4.6$\,\AA.  The 
total exposure time was 2 hours, comprising 3 exposures of 40 minutes each. 
Data reduction was performed in a standard manner using the GEMINI.GMOS 
package within the reduction software IRAF.

Radial velocities were measured by cross-correlation, using the FXCOR task 
within IRAF. FS90 galaxies with known redshift were used as templates. In
the case of galaxies showing bright emission lines, the measurement of
radial velocities was made by fitting these lines (Fig.\,2).

\section{Preliminary results}

The main result is the determination of radial velocities for eight FS90
galaxies without previously known redshifts, and for five unclassified 
faint objects. These results let us extend the colour-magnitude relation 
presented by Smith Castelli et al. (2008a) by 4 magnitudes towards its faint 
end (Smith Castelli et al. in preparation, Smith Castelli et al. this Bulletin).
Among the spectroscopic sample, we have also identified three objects that 
could be classified as blue compact dwarf (BCD) galaxies, one from the FS90 
catalogue and two new ones. 

The new data increase the number of galaxies with radial velocity measurements
in the central region of Antlia, from 31 (Smith Castelli et al. 2008a) to 43,
thus representing a 38\% increase. The mean velocity of the galaxies in the 
central field is 2662 km s$^{-1}$ with $\sigma=$ 664 km s$^{-1}$. These values
have to be compared with those determined by Hopp \& Materne (1985) of
2718 km s$^{-1}$ with $\sigma=$ 469 km s$^{-1}$. A more detailed analysis will 
be presented in a forthcoming paper.

\begin{figure}[h]
  \centering
    \includegraphics[width=0.8\textwidth]{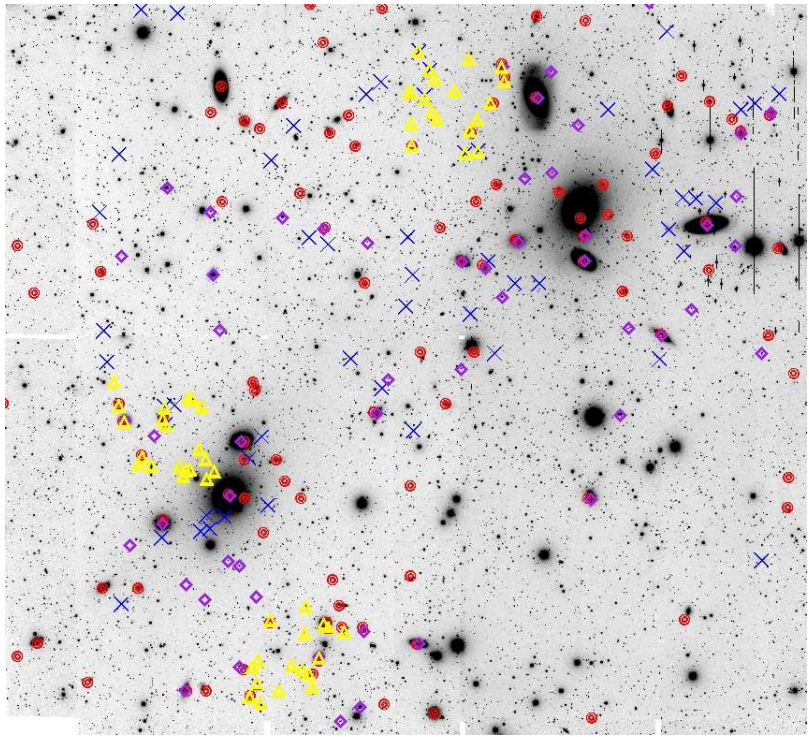}~\hfill%
  \caption{$C+R$ combined image of the central region of Antlia ($36\times36$ 
  arcmin$^2$). North is up and east to
the left. We show the location of FS90 galaxies (circles), new dE and 
dSph galaxy candidates (crosses), MAGELLAN-IMACS targets (diamonds) and 
GEMINI-GMOS targets (triangles). }
  \label{fig1}
\end{figure}

\begin{figure}[h]
  \centering
    \includegraphics[width=0.8\textwidth]{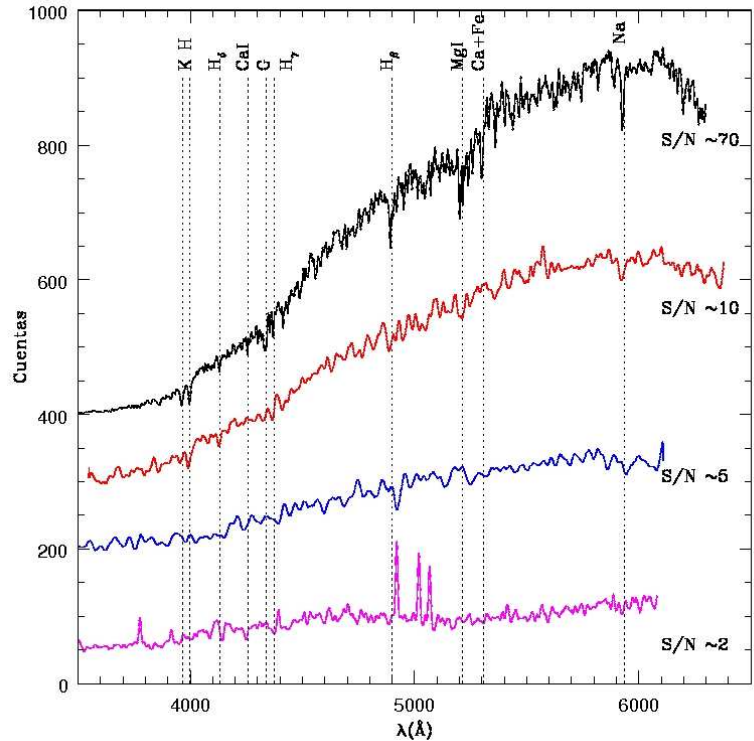}~\hfill%
  \caption{Examples of GEMINI-GMOS spectra with different signal-to-noise 
  ratios.
  From top to bottom the spectra correspond to FS90 133, two newly identified 
Antlia members and FS90 137. The radial velocity of the very faint object 
(FS90 137) was obtained through the emission lines identified in its spectrum. The 
spectra are not flux calibrated and are arbitrarily displaced along the vertical direction 
for clarity.}
  \label{fig2}
\end{figure}

\acknowledgements
Based on observations obtained at the Gemini Observatory, which is operated by 
the AURA, Inc., under a 
cooperative agreement with the NSF on behalf of the Gemini partnership: the 
NSF (United States), the Science and Technology 
Facilities Council (United Kingdom), the National Research Council (Canada), 
CONICYT (Chile), the Australian Research Council (Australia), Ministério da 
Ciência e Tecnologia (Brazil) and SECYT (Argentina). This work was supported by 
grants from CONICET, ANPCyT and UNLP, Argentina.

\begin{referencias}
\vskip 0.5cm

\reference Bassino L. P., Richtler T., Dirsch B., 2008, MNRAS, 386, 1145
\reference Dirsch B., Richtler T., Bassino L. P., 2003, A\&A, 408, 929
\reference Ferguson, H.C. \& Sandage, A. 1990, \aj, 100, 1
\reference Hopp U., Materne J., 1985, A\&AS, 61, 93
\reference Nakazawa, K, Makishima, K., Fukazawa, Y. \& Tamura T. 2000, \pasj, 52, 623
\reference Pedersen, K., Yoshii, Y. \& Sommer-Larsen, J. 1997, \apj, 485, L17
\reference Smith Castelli, A., Bassino, L., Richtler, T., et al. 2008a, MNRAS, 386, 2311  
\reference Smith Castelli A. V., Faifer F. R., Richtler T., Bassino L. P., 2008b, MNRAS, 391, 685
\end{referencias}

\end{document}